\documentclass[11pt,a4paper]{article}
\pdfoutput=1
\usepackage{jcappub}

\usepackage{epsfig,verbatim}
\usepackage{amsmath, amssymb, graphics}
\usepackage{color}
\usepackage{slashed}            
\usepackage{bbm}                
\usepackage{units}              
\usepackage{xspace}             
\usepackage{enumerate}          
\usepackage{soul}
\usepackage{epstopdf}
\usepackage{caption}
\usepackage{subcaption}

\newcommand{\openone}{\leavevmode\hbox{\small1\normalsize\kern-.33em1}}

\usepackage{tensor}

\newcommand{\mathsym}[1]{{}}

\renewcommand\({\left(}
\renewcommand\){\right)}

\newcommand{\ba}{\begin{eqnarray}}
\newcommand{\ea}{\end{eqnarray}}
\newcommand{\be}{\begin{equation}}
\newcommand{\ee}{\end{equation}}

\title{On the viability of \boldmath $m^2\phi^2$ and natural inflation}

\author[a,b]{Ana Ach\'ucarro,}
\author[a]{Vicente Atal}
\author[c]{and Yvette Welling}

\emailAdd{achucar@lorentz.leidenuniv.nl}
\emailAdd{atal@lorentz.leidenuniv.nl}
\emailAdd{welling@strw.leidenuniv.nl}

\affiliation[a]{Instituut-Lorentz for Theoretical Physics, Universiteit Leiden, 2333 CA Leiden, The Netherlands}
\affiliation[b]{Department of Theoretical Physics, University of the Basque Country, 48080 Bilbao, Spain}
\affiliation[c]{Leiden Observatory, Universiteit Leiden, 2300 RA Leiden, The Netherlands}

\abstract{In the context of single field inflation, models with a quadratic potential and models with a natural potential with subplanckian decay constant are in tension with the Planck data. We show that, when embedded in a two-field model with an additional super massive field, they can become consistent with observations. Our results follow if the inflaton is the phase of a complex field (or an angular variable) protected by a mildly broken $U(1)$ symmetry, and the radial component, whose mass is much greater than the Hubble scale, is stabilized at subplanckian values. The presence of the super massive field, besides modifying the effective single field potential, causes a reduction in the speed of sound of the inflaton fluctuations, which drives the prediction for the primordial spectrum towards the allowed experimental values. We discuss these effects also for the linear potential, and show that this model increases its agreement with data as well.}

\begin{document}
\maketitle 

\section{Introduction}

Precise observations of the cosmic microwave background  and large scale structure allow today for a very accurate determination of the cosmological parameters \cite{Adam:2015rua}. This requires the theoretical predictions to be precise and robust, so that there are no uncertainties when interpreting the data, or, more realistically speaking, that any uncertainty is well understood. 
In the particular case of inflation \cite{Guth:1980zm,Starobinsky:1979ty}, the fact that this period might be driven by a single light scalar field, effectively uncoupled to any additional degree of freedom (at least during the time when observable perturbations are generated) is a very appealing scenario in terms of predictability\footnote{At least if we ignore the issue of eternal inflation and the multiverse paradigm.}. Indeed, any given potential has unique and precise predictions, with the only ambiguity coming from uncertainties on which e-foldings correspond to the observable scales. The fact that many models of inflation have been ruled out by measurements of the spectral tilt and bounds on the amount of tensor perturbation, reaffirms this claim. In particular some of the simplest large field models like $m^2\phi^2$ \cite{Linde:1983gd} and natural inflation \cite{Freese:1990rb} are in tension with the data \cite{Ade:2015lrj}. In this paper we show that a particular class of two-field 
embeddings, where the additional field is super heavy, can bring these models back into consistency with the data, by changing the value of the slow roll parameters as well as by generating a reduced speed of sound $c_s<1$ for the fluctuations. We emphasize that this deformation is coming entirely from super massive degrees of freedom (masses much heavier than the Hubble parameter), and is an example of how  important heavy fields may be in determining the low energy effective description. 

In general, a heavy field may influence the low energy dynamics by either affecting the background, the perturbations or a combination of both. We would like to stress that this does not require any high energy excitations or particle production, and our discussion in this paper focuses on the regime in which they do not occur. First of all, changes in the background may come when  evaluating the action at the vacuum expectation value (v.e.v.) of the heavy field. Provided the kinetic energy is dominated by the light field, it is possible to write a Lagrangian in which we recover single-field inflation described by some effective potential (see e.g. \cite{Rubin,Dong:2010in}). Changing the background will result in different  values for the slow roll parameters $\epsilon$ and $\eta$, and thus we call this the \emph{background} model. Secondly, for certain (derivative) couplings between the light and heavy field, perturbations of the heavy field contribute to the low frequency mode and therefore to the low 
energy - 
single field - effective field theory (EFT) for the perturbations\footnote{The 
isocurvature 
modes are heavy and decay fast, they do not 
source the curvature perturbations after horizon crossing. This situation is different from so-called multifield inflation, where there are multiple light fields \cite{Gordon:2000hv,
GrootNibbelink:2001qt}, and quasi-single
field inflation, where the mass of the heavy field is order of the Hubble parameter \cite{Chen:2009we}.}. The effect of this coupling has been widely studied, both in the cases 
when it is small \cite{Pi:2012gf,Chen:2012ge}, and the cases in which it is large \cite{Tolley:2009fg,Achucarro:2010jv,Peterson:2010np,Cremonini:2010ua,Achucarro:2010da,Baumann:2011su,Shiu:2011qw,Avgoustidis:2012yc,Achucarro:2012yr,Burgess:2012dz,Gao:2012uq,Cespedes:2013rda,Baumann:2015nta} (see \cite{Achucarro:2015bra} for a comparative study of some of these works). In the latter case, integrating out the heavy field results in a reduced speed of sound for the adiabatic fluctuations. This is a purely quantum effect that arises when considering the full two-field evolution for the perturbations.

As we will explicitly show through examples, a proper description of the system demands taking both 
effects into account.
In order to do so, we embed inflation in a simple two-field realization such that inflation takes place on a turning trajectory whose radius of curvature is changing very slowly. In our embedding, the inflaton is the phase of a complex field where the $U(1)$ symmetry is mildly broken and the v.e.v. of the - massive - radial field is approximately the radius of curvature of the inflationary trajectory (see \cite{McDonald:2014oza,Barenboim:2014vea} and \cite{Li:2014vpa} for realizations in field theory and supergravity respectively). 

As already anticipated, we will show that within this framework quadratic inflation can be consistent with the data, and that subplanckian values of the effective (instantaneous) decay constant are no longer disfavored in natural inflation. This will require that the v.e.v. of the radial field takes subplanckian values.
Different deformations of standard natural and chaotic inflation were discussed in e.g. \cite{Kim:2004rp,Dimopoulos:2005ac,Kappl:2015pxa,Li:2015mwa}, and \cite{Pallis2010287,Kallosh:2013tua,Ashoorioon:2013eia,Kannike:2015apa,Li:2015mwa,Boubekeur:2015xza,Buchmuller:2015oma} respectively. Unlike previous studies that included additional heavy fields to improve agreement with the data due to a flattening of the potential, in all cases described here the effective potential {\it steepens}. However the speed of sound effects dominate and move the predictions {\it downward} in the $(n_s,r)$ plane, towards the best fit region. In addition, in some cases the speed of sound slowly decreases along the trajectory, causing a shift towards higher vaues of $n_s$.

The outline of the paper is as follows. In section \ref{secgen} we introduce a simple two-field embedding of inflation with an additional heavy field. We explain how both the background and the curvature perturbations are affected by the presence of the heavy field. Furthermore we provide analytical expressions for the observables. In sections \ref{secquad}, \ref{seclin} and \ref{secnat} we study the predictions of the observables of quadratic, linear and natural inflation embedded in this two-field scenario. We show explicitly how the predictions in the $(n_s,r)$ plane move towards the $1\sigma$ allowed region of Planck. Finally in section \ref{seccon} we discuss our findings and conclude. 
 
\section{General Setup}\label{secgen}

\subsection{Two-field embedding}
We embed inflation in a simple two field realization, given by the following Lagrangian:
\be\label{eq:full}
\mathcal{L}=\frac{1}{2}\partial_{\nu}\rho\partial^{\nu}\rho+\frac{1}{2}\rho^2\partial_{\nu}\theta\partial^{\nu}\theta-\frac{m_{\rho}^2}{2}\(\rho-\rho_0\)^2-V(\theta).
\ee
Here we assume the inflaton to be the phase of a complex field, and the $U(1)$ symmetry has been mildly broken by a potential $V(\theta)$. This model was already studied in \cite{Cremonini:2010ua} (with a different kinetic term), and in \cite{Achucarro:2012yr}, for the case of linear inflation (that we discuss below). This is not the most general Lagrangian consistent with the symmetries invoked. In fact we may also have different choices for the potential or the field space metric, but we choose this form as the simplest starting case. We also notice that since the model becomes singular at $\rho=0$, additional degrees of freedom should appear at some higher energy scale. Additionally, these models will have  monodromy; the potential is not completely invariant after the phase $\theta$ has made a $2\pi$ cycle 
\cite{McAllister:2008hb,Kaloper:2008fb,Kaloper:2011jz,Harigaya:2014eta}.

The standard logic (that we will show is inaccurate) is that, if the radial field is sufficiently heavy, the field $\rho$ will rapidly reach its minimum at $\rho_0$, so that one may truncate the model and consider the single field Lagrangian 
\be
\mathcal{L}=\frac{1}{2}\rho_0^2\partial_{\nu}\theta\partial^{\nu}\theta-V\(\theta\), \\
\ee

\noindent which by a field redefinition becomes:
\be
\mathcal{L}=\frac{1}{2}\partial_{\nu}\phi\partial^{\nu}\phi-V\(\phi/\rho_0\). \\
\ee

We will show that truncating the model in this manner yields inaccurate predictions. The reason is twofold. First, because of the kinetic coupling the radial field will have a minimum at $\bar{\rho}\neq\rho_0$. Plugging this solution back in the Lagrangian will result in an EFT in which we recover single-field inflation described by an effective potential \cite{Rubin,Dong:2010in}. In general, this will result in different  predictions for both $\epsilon$ and $\eta$ (the slow roll parameters) at the observable scales. This single field description is possible provided the kinetic energy is dominated by the angular field, or more specifically that $\dot{{\rho}}^2+{\rho}^2\dot{\theta}^2=\(\(d{\rho}/d\theta\)^2+{\rho}^2\)\dot{\theta}^2\sim{\rho}^2\dot{\theta}^2$. From this condition we will demand that $d{\rho}/d\theta\ll{\rho}$. Secondly, light and heavy field perturbations will be coupled through the angular velocity, $\dot{\theta}/H$, which, if large, will give rise to a low energy EFT with a reduced speed of 
sound 
$c_s$ for the adiabatic fluctuations \cite{Tolley:2009fg,Achucarro:2012sm,Achucarro:2012yr,Gao:2012uq}.

The prediction for the system can then be computed with the usual relations\footnote{We actually use the more precise predictions for r as in \cite{Palma:2014faa}, where the difference in freeze-out time between the scalar and tensor perturbation is considered \cite{Baumann:2014cja}. This effect becomes relevant when $c_s\ll1$. We also compute the power spectrum, from which we derive $n_s$, at second order in slow roll, as in \cite{Palma:2014faa}.} $r=16\epsilon c_s$ and $n_s=1-2\epsilon-\eta-s$ with $\epsilon\equiv-\dot{H}/H^2$,  $\eta\equiv\dot{\epsilon}/H\epsilon$ and $s\equiv{\dot{c_s}}/{Hc_s}$. From these expressions it is clear that a reduced speed of sound, $c_s<1$, will contribute to moving the predictions of the model towards smaller values of $r$ (taking into account that in general $\epsilon$ will also change while we go from $c_s=1$ to $c_s\ll1$). Additionally,   $n_s$ will also change making theories flow in the $(n_s,r)$ plane\footnote{In \cite{Zavala:2014bda}, the effect of $c_s$ and $s$ in 
the 
$(n_s,r)$ plane for a phenomenological ansatz $c_s(N)$ was discussed.}. In the following, we show how to exactly calculate these quantities.

\subsection{Analytical predictions}
The possibility to make analytical predictions for $n_s$ and $r$ depends on the ability to calculate the radius $\bar{\rho}(\theta)$ at which the radial field stabilizes, as a function of the parameters of the two-field embedding. While it is not generically possible to solve the full two-field model, there is a regime in which such analytical predictions are possible. This is the regime in which the time derivatives of $\rho$ can be neglected in the equations of motion (e.o.m.). In order to show this explicitly, consider the e.o.m. for the system
\ba
&\ddot{\rho}+3H\dot{\rho}-\rho\dot{\theta}^2+V_{\rho}=0 \\
&\rho^2\ddot{\theta}+2\rho\dot{\rho}\dot{\theta}+3H\rho^2\dot{\theta}+V_{\theta}=0
\ea

\noindent together with the Friedmann equation (from here on we set the reduced Planck mass $m_{pl}\equiv(8\pi)^{-1/2}M_{pl}=1$)
\be
3H^2=\frac{1}{2}\left(\dot{\rho}^2+\rho^2\dot{\theta}^2\right)+V.
\ee
\noindent where $V_{\theta}=\partial V/\partial\theta$ and $V_{\rho}=\partial V/\partial\rho$.

 First, we assume that we can neglect the derivatives of $\rho$ in the previous equations (which is a good approximation in all of the cases studied here). We assume then $\ddot{\rho},3H\dot{\rho} \ll \rho\dot{\theta}^2,V_{\rho}$ and $2\rho\dot{\rho}\dot{\theta}\ll3H\rho^2\dot{\theta},V_{\theta},\rho^2\ddot{\theta}$ . Let us note that the previous inequalities demand that, at the same time, $\frac{1}{ \bar{\rho}}\frac{d\bar{\rho}}{d\theta} \ll \frac{\dot{\theta}}{3H} $ and $\frac{1}{ \bar{\rho}}\frac{d\bar{\rho}}{d\theta} \ll \frac{3H}{2\dot{\theta}}$. This directly implies that
\be
\frac{1}{ \bar{\rho}}\frac{d\bar{\rho}}{d\theta}\ll 1
\ee
\noindent which is the condition for writing a single field model for the background (i.e. that the kinetic energy is dominated by the angular velocity). This condition does not mean that the field $\rho$ has to be exactly constant, but rather that its time evolution is slow.\\
Furthermore, as $\theta$ plays - mainly - the role of the inflaton, we also drop $\ddot{\theta}$. This demands $\ddot{\theta}<3H\dot{\theta}$. The simplified system then reads:
\ba
&\rho\dot{\theta}^2=V_{\rho} \label{eq:sim1}\\
&3H\rho^2\dot{\theta}+V_{\theta}=0 \label{eq:sim2}
\ea
and 
\be\label{eq:sim3}
3H^2=V .
\ee

Importantly, let us note that there is no bound on $\dot{\theta}/H$ (as long as $\dot{\theta}<m_{\rho}<M_{pl}$). This quantity  plays an important role in determining both the coupling of the perturbations between the light and heavy field, and - as we will see below - the slow roll parameters.\\

In principle with these equations we obtain $\bar{\rho}(\theta)$, plug the solution in the potential and find a canonical variable so that we have a single-field effective potential. However, in situations in which the solution $\bar{\rho}(\theta)$ is a complicated function of $\theta$, it may be too difficult to follow this procedure, the main reason being that we need to find a canonical variable $\phi$ such that $\bar{\rho}\dot{\theta}=\dot{\phi}$. Nonetheless, the system can still be solved semi-analytically in a single field approach. If the kinetic energy is dominated by $\theta$, then $\epsilon$ is given by
\be
\epsilon=\frac{1}{2\bar{\rho}^2}\(\frac{V_{\theta}}{V}\)^2 .
\ee
With this we can calculate $\eta=\dot{\epsilon}/\epsilon H$ giving
\be
\eta=\frac{2}{\bar{\rho}^2}\(\frac{V_{\theta\theta}}{V}\)+4\epsilon-2\frac{\dot{\bar{\rho}}}{\bar{\rho} H} .
\ee

\noindent Importantly, the last term cannot be neglected. Indeed
\ba
\delta&\equiv&\frac{\dot{\bar{\rho}}}{\bar{\rho} H}\\
&=&\frac{1}{\bar{\rho}}\frac{d\bar{\rho}}{d\theta}\frac{\dot{\theta}}{H}=\frac{1}{\bar{\rho}}\frac{d\bar{\rho}}{d\theta}\frac{\sqrt{2\epsilon}}{\bar{\rho}} .
\ea
While the reduced e.o.m. demands $\delta\ll 1$, $\delta$ may be $\mathcal{O}(\epsilon,\eta)$. Then, we can calculate all of the relevant quantities for the background with the following relations\footnote{These expressions were derived assuming a separable potential. For non separable potentials they remain approximately valid provided $V_{\theta\rho}(d\bar{\rho}/d\theta)\ll V_{\theta\theta}$}:
\be\label{eq:sf_pred}
\epsilon=\frac{1}{2\bar{\rho}^2}\(\frac{V_{\theta}}{V}\)^2\quad,\quad\eta=\frac{2}{\bar{\rho}^2}\(\frac{V_{\theta\theta}}{V}\)+4\epsilon-2\delta\quad,\quad\delta=\frac{\dot{\bar{\rho}}}{\bar{\rho} H}\quad,\quad N=\int \bar{\rho}^2\frac{V}{V_{\theta}}d\theta ,
\ee
\noindent where $\bar{\rho}$ is the solution to $\rho^3 V_\rho=V_\theta^2/3V$ and $N$ is the number of e-folds before the end of inflation.
 From here it is clear that the time dependence of $\bar{\rho}$ has to be explicitly taken into account in order to make accurate predictions. This mean that while we can neglect the derivatives of $\rho$ in the e.o.m. its derivatives do play an important role in determining the observables of the model.

 As for the perturbations, in a regime in which the angular acceleration $\ddot{\theta}$ is small in comparison with the effective mass of the heavy field \cite{Cespedes:2012hu} (given here by $m^2_{\text{eff}}=m_{\rho}^2-\dot{\theta}^2$ and demanding $m_{\text{eff}}\gg H$), the low energy EFT develops a speed of sound of the fluctuations which is given by\footnote{The speed of sound presented here is the $k\rightarrow 0$ limit of the speed of sound obtained by integrating out the heavy mode. The $k$-dependence is not extremely important to compute the observables of the theory - at least for moderate reductions in $c_s$ -  but it becomes important in order to assess the overall consistency and predictivity of the theory \cite{Baumann:2011su,Gwyn:2012mw,Gwyn:2014doa}. Whenever needed, we use the full $k$-dependent $c_s$.}
\be\label{eq:cs}
c_s^{-2}=1+4\frac{\dot{\theta}^2}{m_{\text{eff}}^2} \quad\quad\quad\text{where}\quad m^2_{\text{eff}}=m_{\rho}^2-\dot{\theta}^2 .
\ee
We refer to \cite{Achucarro:2012yr} for a more detailed discussion.
Moreover, it is easy to show that $s$ can be written solely in terms of $\epsilon$, $\eta$, $\delta$ and $c_s$:
\be\label{eq:s}
s=\(\epsilon-\frac{\eta}{2}+\delta\)\(1-c_s^2\)\(\frac{3}{4}+\frac{1}{4\,c_s^2}\) .
\ee

With all these elements it is possible to compute all of the observables of the model, i.e. $r$ and $n_s$, without having to solve any dynamical equation, as in the standard slow roll computation. As can be seen from eq. (\ref{eq:cs}), in order to have a substantial reduction in the speed of sound we will need large angular velocities, of the order of the effective heavy mass. This is consistent with slow roll whenever the radius of curvature is small enough, and that is the reason why we will demand the condition $\rho_0<1$  to be satisfied. 

Before closing this section, two comments are in order.
First, it is important to ensure that the theory stays weakly coupled up to the scale where new physics cannot be further integrated out. In models with a reduced speed of sound, this places a theoretical lower bound on the speed of sound \cite{Baumann:2011su,Gwyn:2012mw}. Every case presented here is consistent with this bound, provided a scale dependent speed of sound - like the one we have - is taken into account.

Secondly, a reduction in the speed of sound unavoidably implies a cubic interaction for the adiabatic perturbation ~\cite{Cheung:2007st}, producing potentially observable non-gaussianity. In particular, for the case of a nearly constant speed of sound we have \cite{Chen:2006nt}\cite{Achucarro:2012yr}:
\be
f_{NL}^{(eq)}=\frac{125}{108}\frac{\epsilon}{c_s^2}+\frac{5}{81}\frac{c_s^2}{2}\(1-\frac{1}{c_s^2}\)^2+\frac{35}{108}\(1-\frac{1}{c_s^2}\) .
\ee
This means that in order to have a measurable non gaussianity $|f_{NL}^{eq}|>5$, we need $c_s<0.2$ (which is still consistent with weak coupling  \cite{Baumann:2011su,Gwyn:2012mw}).  While it is interesting to search for such values of $c_s$, we will notice that much milder reductions in the speed of sound can already leave big imprints in the power spectrum, and we are thus going to focus  mainly on mild reductions of $c_s$.

\section{Quadratic Inflation}\label{secquad}

\indent Our first example to show how a heavy field may influence the low energy dynamics is a two-field embedding of the quadratic inflation model. The Lagrangian for the single field model \cite{Linde:1983gd} is given by:
\be\label{eq:cao}
\mathcal{L}=\frac{1}{2}\partial_{\nu}\phi\partial^{\nu}\phi-\frac{1}{2}m_{\phi}^2\phi^2 .
\ee
We embed this model in the two field scenario (\ref{eq:full}), and consider the following Lagrangian:
\be\label{eq:full_cao}
\mathcal{L}=\frac{1}{2}\partial_{\nu}\rho\partial^{\nu}\rho+\frac{1}{2}\rho^2\partial_{\nu}\theta\partial^{\nu}\theta-\frac{m_{\rho}^2}{2}\(\rho-\rho_0\)^2-\Lambda^4\theta^2 .
\ee
Assuming $\rho=\rho_0$, and defining $\phi=\rho_0\theta$, we recover the single field Lagrangian with the mass $m_{\phi}$ given by $m_{\phi}^2=2\Lambda^4/\rho_0^2$. Thus, at the level of this truncation, both Lagrangians ($\ref{eq:cao}$) and  ($\ref{eq:full_cao}$) are equivalent. Going beyond this simplification demands solving the full e.o.m. Fortunately we can rely on the reduced e.o.m. to find approximate solutions.
Solving equations (\ref{eq:sim1}),(\ref{eq:sim2}) and (\ref{eq:sim3}), the minimum in the radial direction, $\bar{\rho}$, is given by the root of the following equation:
\be\label{eq:rootCao}
\bar{\rho}^3\(\bar{\rho}-\rho_0\)-\frac{4}{3}\frac{\Lambda^4}{m_{\rho}^2}=0,
\ee
\noindent while the angular velocity is given by
\be
\dot{\theta}=-\frac{2}{\sqrt{3}}\frac{\Lambda^2}{\bar{\rho}^2} .
\ee
Here we have used $V=V(\theta)$, which, as we will remark below, is a very good approximation. With these solutions at hand we can then predict how the observables move in the $(n_s,r)$ plane. In doing so we will split the effects on the background and perturbations.\\

\noindent \emph{Background model:}  All the relevant quantities for calculating the background can be found in equation (\ref{eq:sf_pred}). First of all, because $\bar{\rho}\neq\rho_0$, the potential $V$ will have a contribution of the form $V_0=\frac{m_{\rho}^2}{2}\(\bar{\rho}-\rho_0\)^2$. It is easy to show  that this contribution is negligible in comparison with $V(\theta)$ in the computation of the slow roll parameters at $N=50-60$\footnote{This approximation  breaks down towards the end of inflation. The numerical results confirm that it does not affect the predictions at the observables scales.}. Thus, we can use $V\sim V\(\theta\)$. Under this simplification, and because $\bar{\rho}\sim cte$, the background model yields the same predictions as in the standard quadratic inflation, i.e. $\epsilon=1/2 N$ and $\eta=1/N$.\\

\begin{figure}[h!]
\begin{subfigure}{.6\textwidth}
  \centering
  \includegraphics[width=1\linewidth]{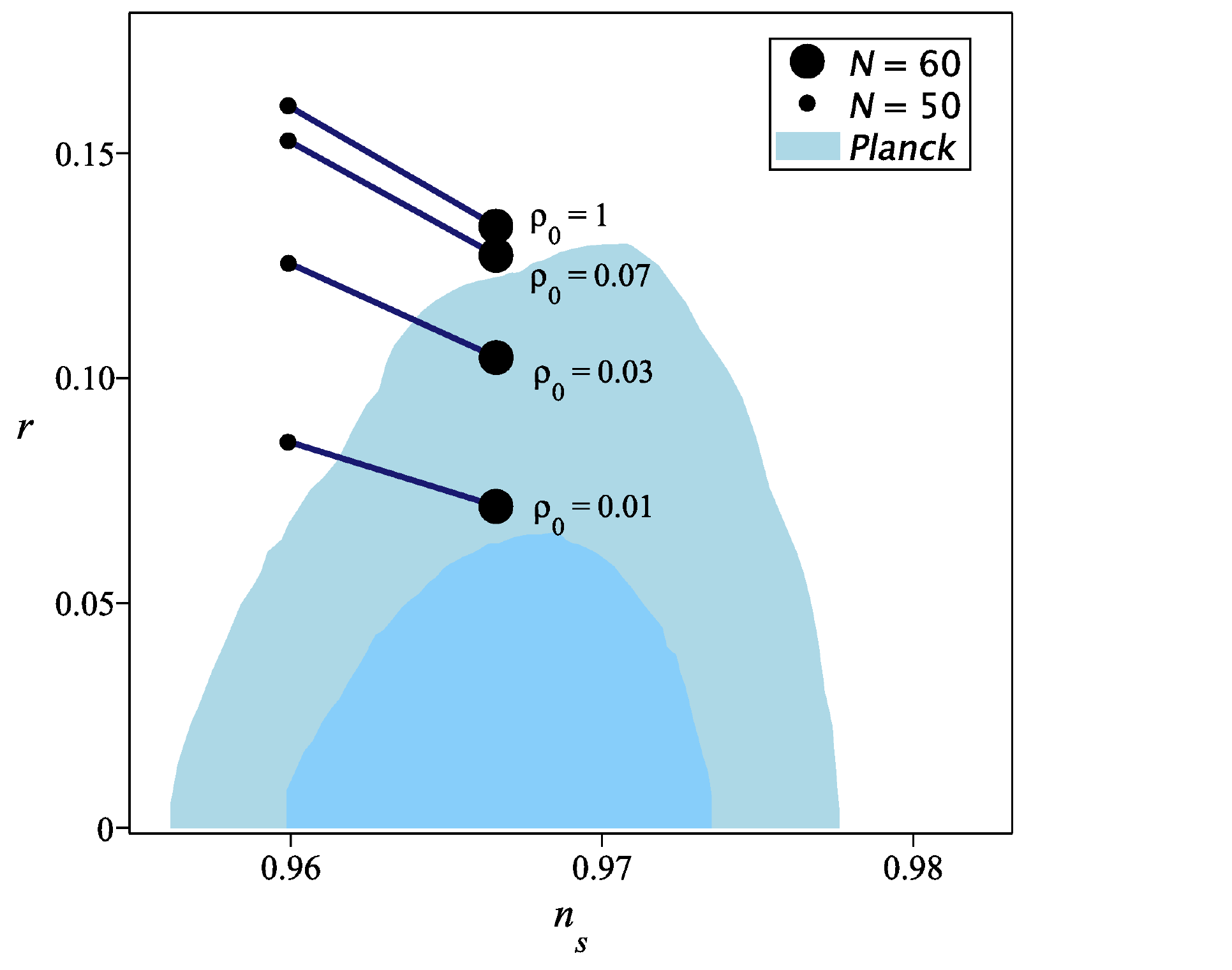}
  \label{fig:sub1}
\end{subfigure}%
\begin{subfigure}{.4\textwidth}
  \centering
  \includegraphics[width=1\linewidth]{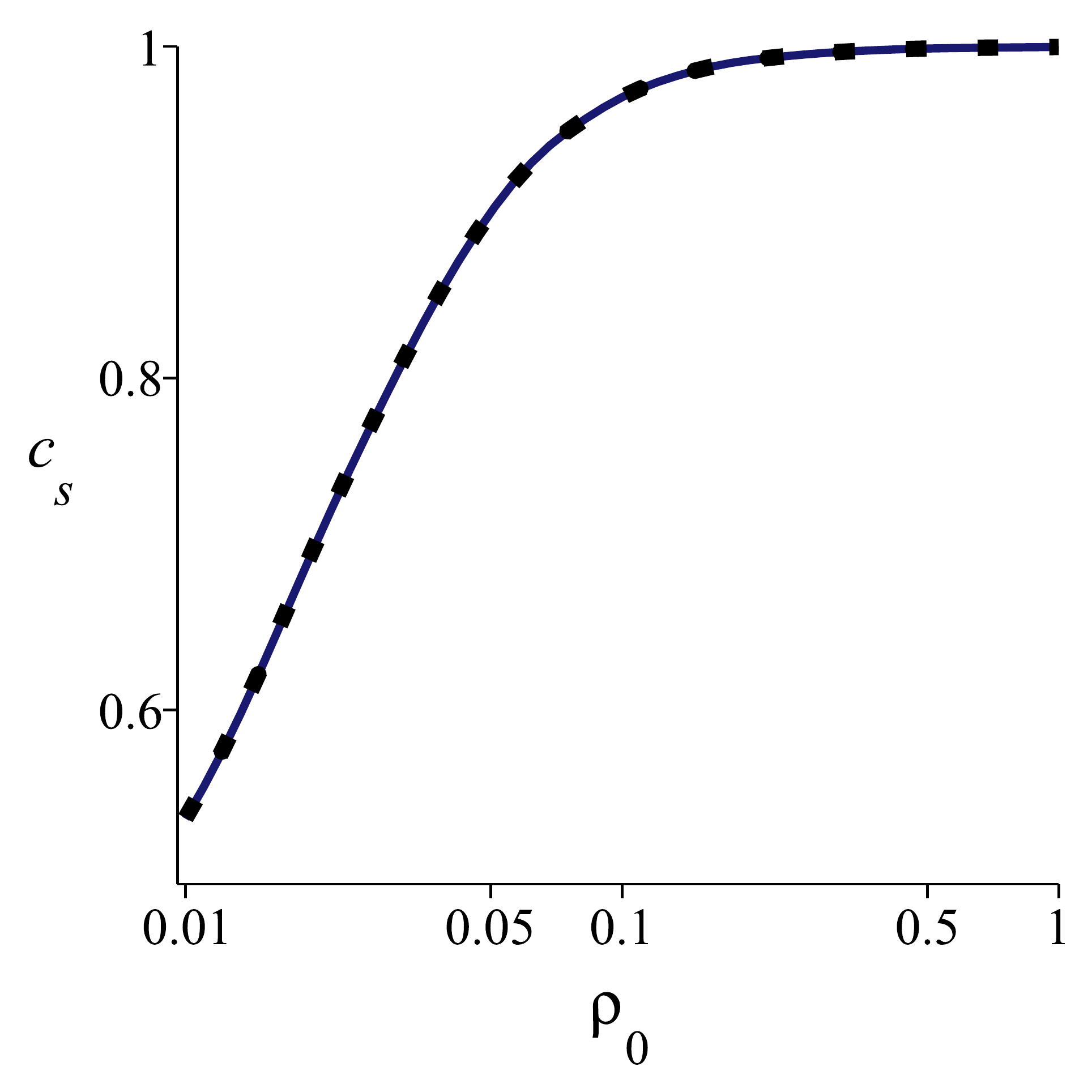}
  \label{fig:cao}
\end{subfigure}
\caption{\emph{Left}: The $(n_s, r)$ plane for $m^2\phi^2$ inflation when embedded in the model given by ($\ref{eq:full_cao}$), with the mass of the heavy field given by $m_{\text{eff}}^2=100H^2$. The predictions are calculated using the EFT, which is an excellent description fo the full two-field system. The blue regions are the 1-$\sigma$ and 2-$\sigma$ allowed regions from Planck \cite{Ade:2015lrj}. \emph{Right}: Speed of sound of the adiabatic fluctuation, given by equation (\ref{eq:cs}) for $N=50-60$. The dotted line is computed using the numerical solution of the full two-field system ($\ref{eq:full_cao}$), and the solid line is computed with the semi-analytical approximation of equation (\ref{eq:sf_pred}).}
\label{fig:csPhisq}
\end{figure}

\noindent \emph{EFT for the perturbations}: While the background does not change as we change $\rho_0$, we find that perturbations develop a constant speed of sound which is noticeably different from 1 for values of $\rho_0<0.1$, as can be seen in the right panel of figure \ref{fig:csPhisq}.\\

Putting all these elements together we compute the prediction for $(n_s,r)$. Since $\epsilon$ and $\eta$ are unchanged, and $s\sim0$, only the tensor to scalar ratio is going to be modified, and its modification will only be due to the change in $c_s$. We test these predictions with a numerical solution of the two-field system (partly done using the code from \cite{Dias:2015rca}), choosing $\rho_0$ ranging from $0.01$ to $1$ and $m_{\rho}$ such that $\(m_{\text{eff}}/H\)^2=100$ in the observable scales. We fix $\Lambda$ such that we have the 
right amplitude for the perturbations. Let us note that we have fixed the effective mass of the heavy field (which is always smaller than the bare mass $m_{\rho}$) such that it is much greater than the Hubble parameter. Our results are summarized in figure \ref{fig:csPhisq}, displayed together with the experimental bounds from Planck\footnote{From the PLA-PR2-2015 official chains including TT,TE and low-$l$ polarization data  at \url{http://pla.esac.int/pla/}.} \cite{Ade:2015lrj}. First of all, there is very good agreement between the predictions of the analytical single field EFT and the full two-field system, and more importantly, there are sizeable effects in terms of where the predictions lie in the $(n_s,r)$ plane.

In particular, we see that  when decreasing $\rho_0$ the quadratic potential becomes more consistent with Planck confidence regions. This is not a surprising result, since we know that reduced speeds of sound lead to smaller values for the tensor to scalar ratio, but the fact that this can be achieved with the simple quadratic potential ``UV completed''   with an additional very massive field is worth noting. Let us note that larger reductions in the speed of sound can easily be attainable provided we consider smaller values for $\rho_0$. This will further increase the consistency with the data as well as generating potentially observable non gaussianities (e.g for, $\rho_0=10^{-3}$ we find $|f_{NL}^{eq}|\sim5$).

\section{Linear Inflation}\label{seclin}

We repeat the analysis, this time for the linear inflaton potential \cite{McAllister:2008hb}. We consider the following Lagrangian:
\be\label{eq:full_lin}
\mathcal{L}=\frac{1}{2}\partial_{\nu}\rho\partial^{\nu}\rho+\frac{1}{2}\rho^2\partial_{\nu}\theta\partial^{\nu}\theta-\frac{m_{\rho}^2}{2}\(\rho-\rho_0\)^2-\alpha\theta 
\ee
This embedding was already studied in \cite{Achucarro:2012yr}, where very small values of $\rho_0$ ($\rho_0\sim10^{-4}$) were considered in order to find large reductions in the speed of sound, as a working example of how decoupling works in this setup. We complement those results with the predictions for the values of $\rho_0$ considered here, in order to show how the theory flows from the vanilla linear potential to the new predictions. The v.e.v. of the radial field can be found by solving the reduced e.o.m.. Again, we can set $V=V(\theta)$, so that we have the following algebraic equation for the radial field $\bar{\rho}$:
\be\label{eq:algLin}
\frac{\alpha}{3\bar{\rho}^4\theta}=m_{\rho}r^2\(1-\frac{\rho_0}{\bar{\rho}}\)
\ee

\begin{figure}[h!]
\centering
\begin{subfigure}{.6\textwidth}
  \centering
  \includegraphics[width=1\linewidth]{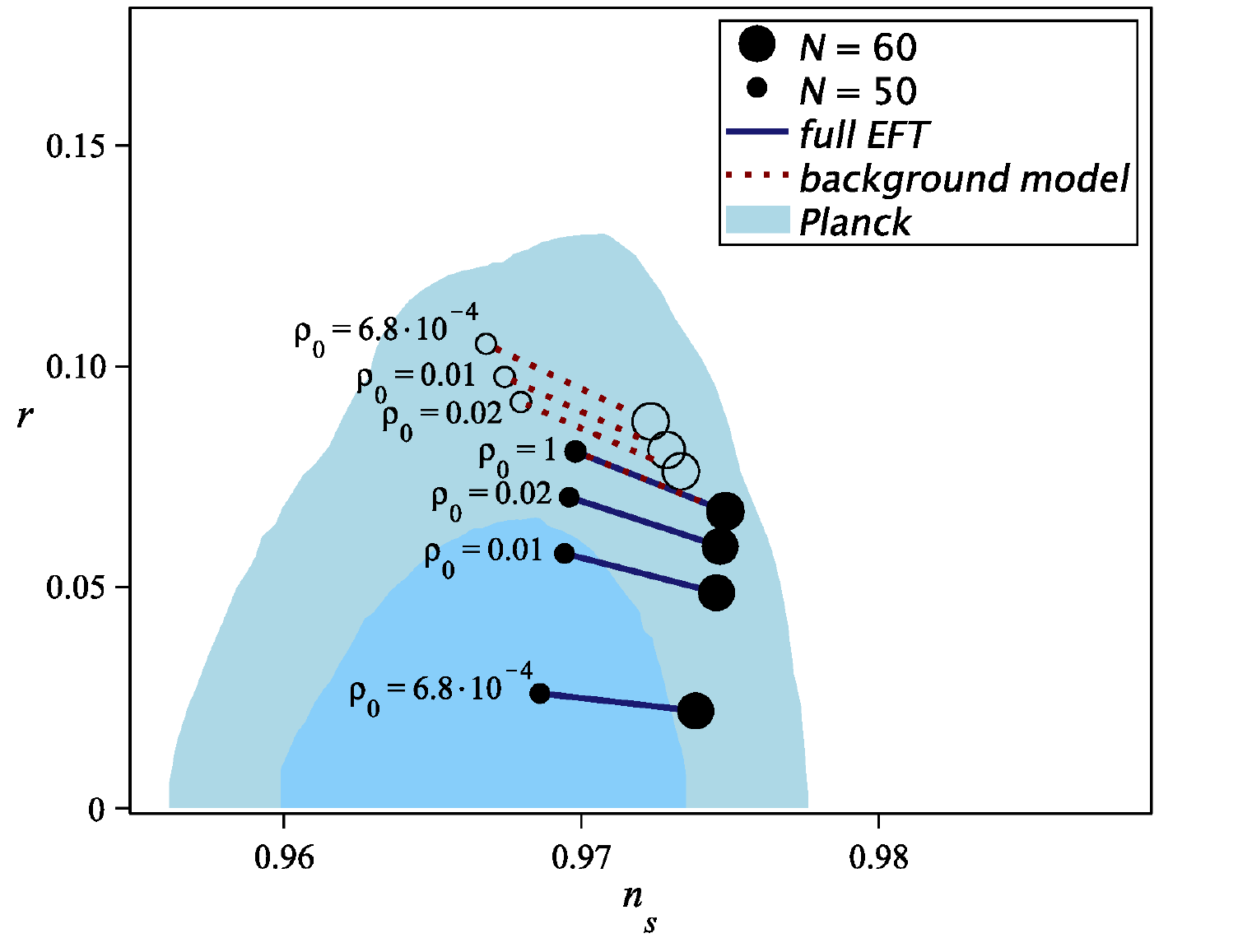}
  \label{fig:sub1}
\end{subfigure}%
\begin{subfigure}{.4\textwidth}
  \centering
  \includegraphics[width=\linewidth]{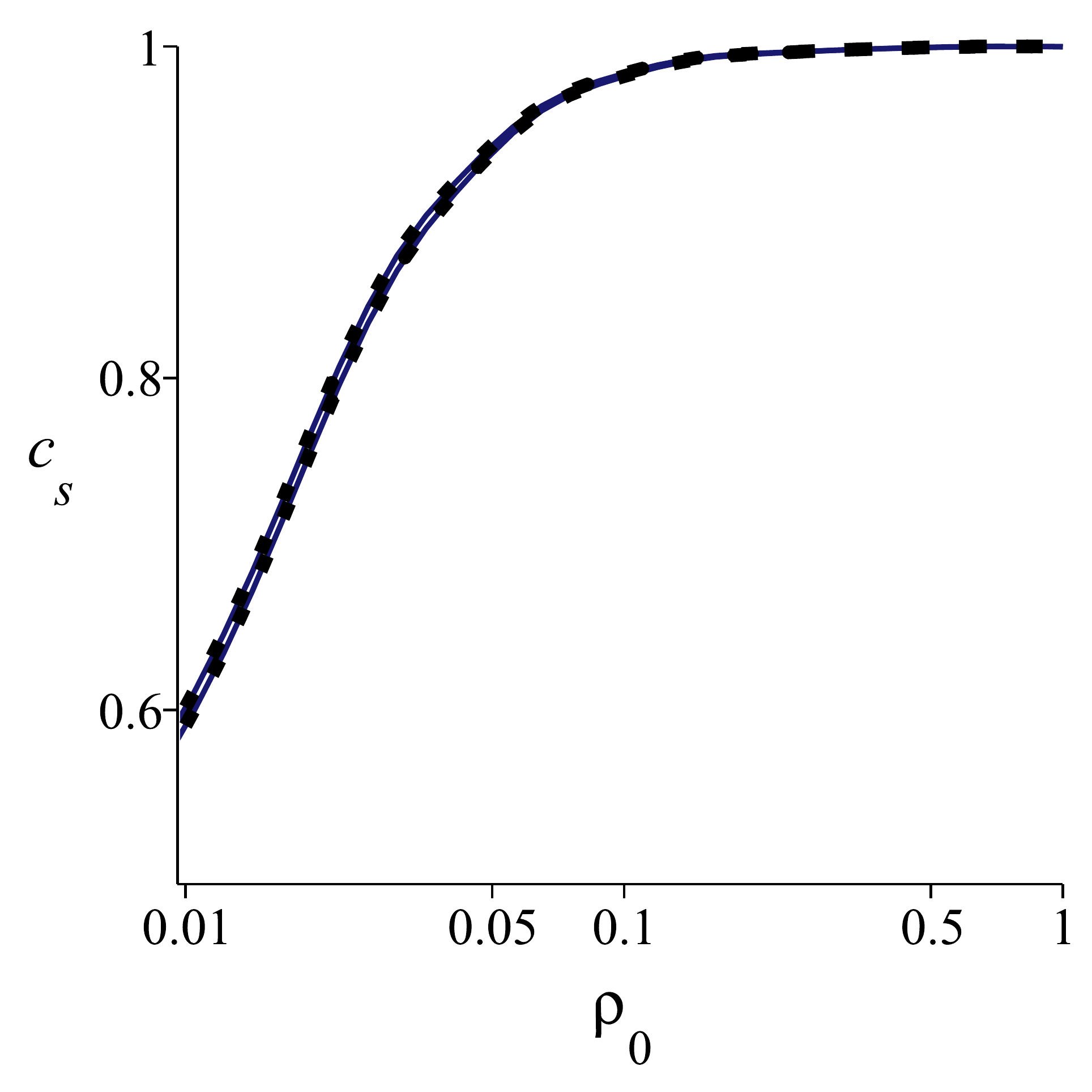}
  \label{fig:sub2}
\end{subfigure}
\caption{\emph{Left}: The $(n_s, r)$ plane for inflation with linear potential when embedded in the two-field model given by ($\ref{eq:full_lin}$), with the mass of the heavy field given by $m_{\text{eff}}^2=100H^2$. The points are calculated using the EFT, which we have checked is an excellent approximation to the numerical two-field prediction. For comparison, we also plot the predictions if we had only considered the effects on the background (red dotted line). \emph{Right}: Speed of sound of the adiabatic fluctuation equation (\ref{eq:cs}) for $N=50-60$. The dotted line is computed using the full two-field numerical solution of (\ref{eq:full_lin}), and the solid line is the semi-analytical single-field approximation equation (\ref{eq:sf_pred}).}
\label{fig:csLin}
\end{figure}

\noindent Here, the solution for $\bar{\rho}$ will explicitly depend on both $\theta$ and $\rho_0$. However, the dynamics in $\rho$ is such that its time derivatives are still negligible in the e.o.m. Because $\bar{\rho}(\theta)$ is a complicated function of $\theta$, it is not easy to find an effective potential. Fortunately, we can solve the system by considering the slow roll parameters as given in (\ref{eq:sf_pred}). The solution of this system is such that $\epsilon$ becomes \emph{bigger} as we decrease $\rho_0$. In principle this is bad news since we would not like to move away from the $1\sigma$ contour of Planck data (see figure \ref{fig:csLin}).  

Fortunately, as $\rho_0$ decreases, $c_s$ decreases, which dominates over the increase in $\epsilon$. This means that the tensor to scalar ratio $r$ decreases as $\rho_0$ decreases. We show these effects in fig. \ref{fig:csLin} where we plot the full EFT i.e. considering  the combined effect of background and perturbations, and the predictions considering only the effects on the background. We also add the prediction for the parameters considered in \cite{Achucarro:2012yr} ($\rho_0=6.8\times10^{-4}$, $m^2_{\text{eff}}=250 H^2$).

As for the quadratic potential, larger reductions in the speed of sound can easily be attainable provided we consider smaller values for $\rho_0$. This will also increase the consistency with the data as well as generating potentially observable non gaussianities (e.g for, $\rho_0=10^{-3}$ we find $|f_{NL}^{eq}|\sim5$).

\section{Natural Inflation}\label{secnat}

\begin{figure}[h!]
\begin{subfigure}{.5\textwidth}
  \includegraphics[width=8.3cm,height=6.8cm]{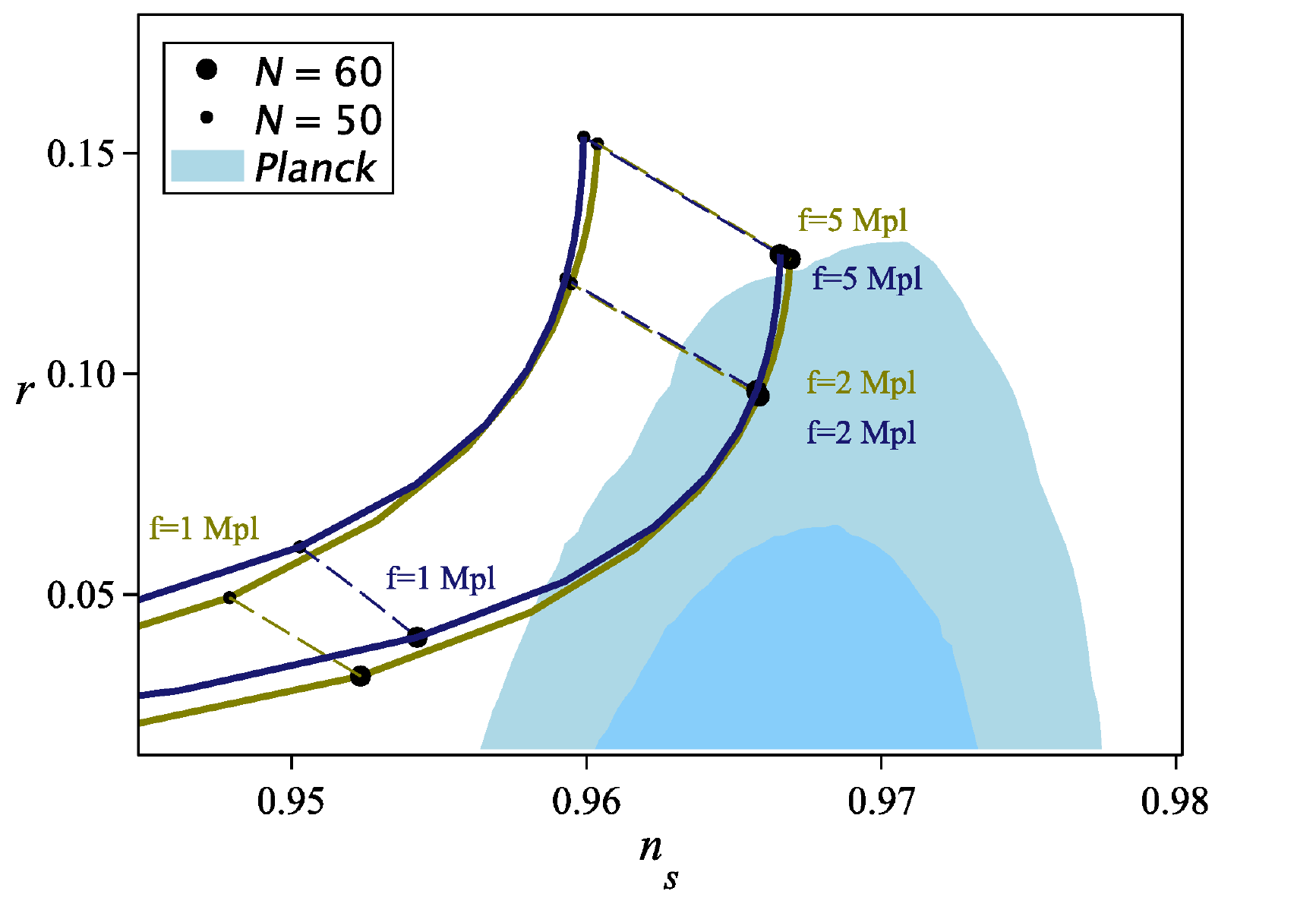}
  \label{fig:sub1}
\end{subfigure}%
\begin{subfigure}{.5\textwidth}
  \includegraphics[width=8.3cm,height=6.8cm]{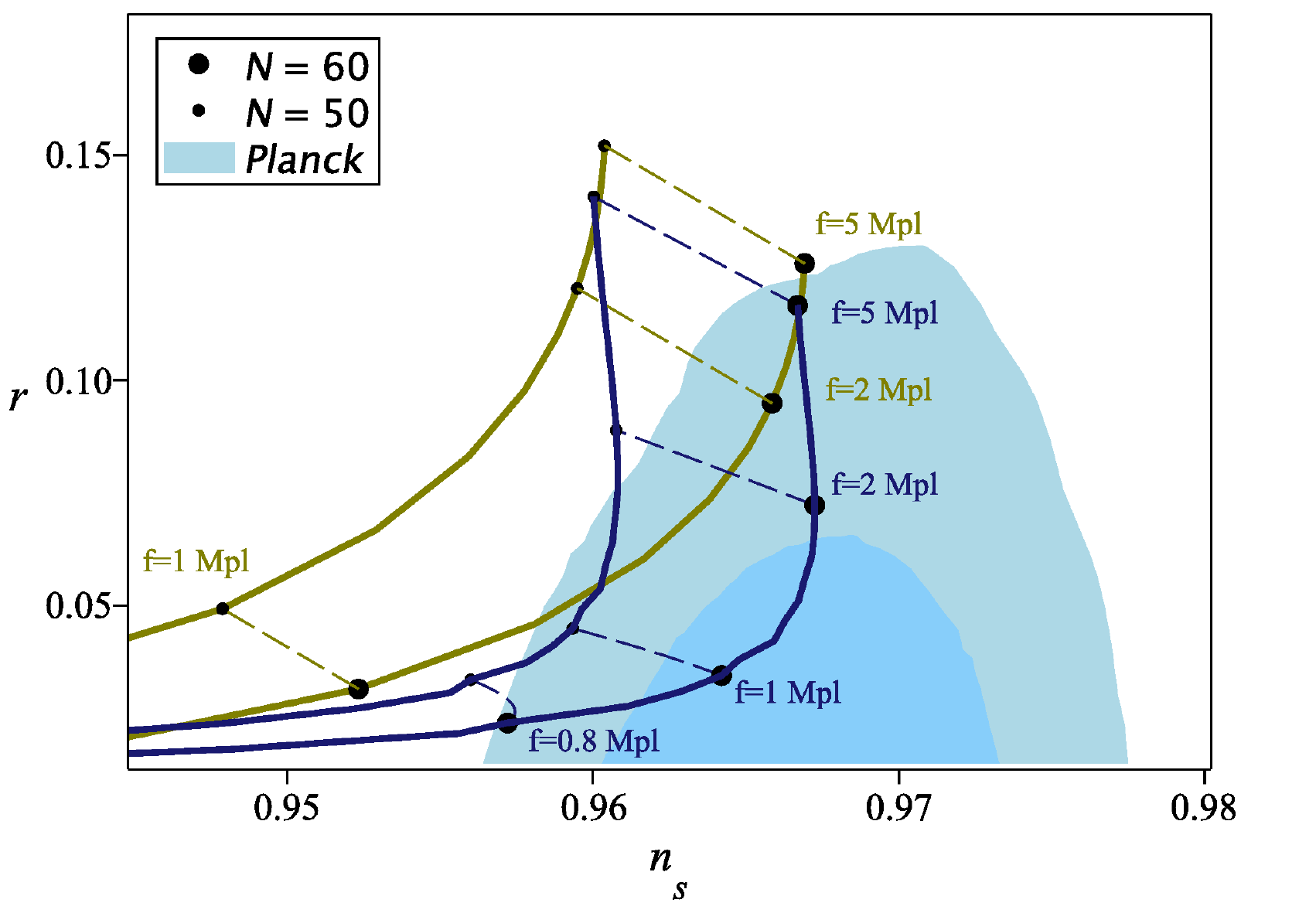}
  \label{fig:sub2}
\end{subfigure}
\caption{The $(n_s, r)$ plane for inflation with the natural potential, with the mass of the heavy field given by $m_{\text{eff}}^2\sim100H^2$. The green line is the standard natural inflation scenario given by (\ref{eq:trunc_nat}). The blue line is the two-field model in (\ref{eq:full_nat}), for different values of the instantaneous decay constant $f=\bar{\rho}/m$. The agreement between the numerical and semi-analytical predictions is excellent. Although the potential steepens, the predictions moves towards the best fit region because of $c_s$. \emph{Left}) $m=0.01$ \emph{Right}) $m=0.002$ }
\label{fig:Nat}
\end{figure}

\begin{figure}[h!]

\centerline{\includegraphics[width=0.7\textwidth]{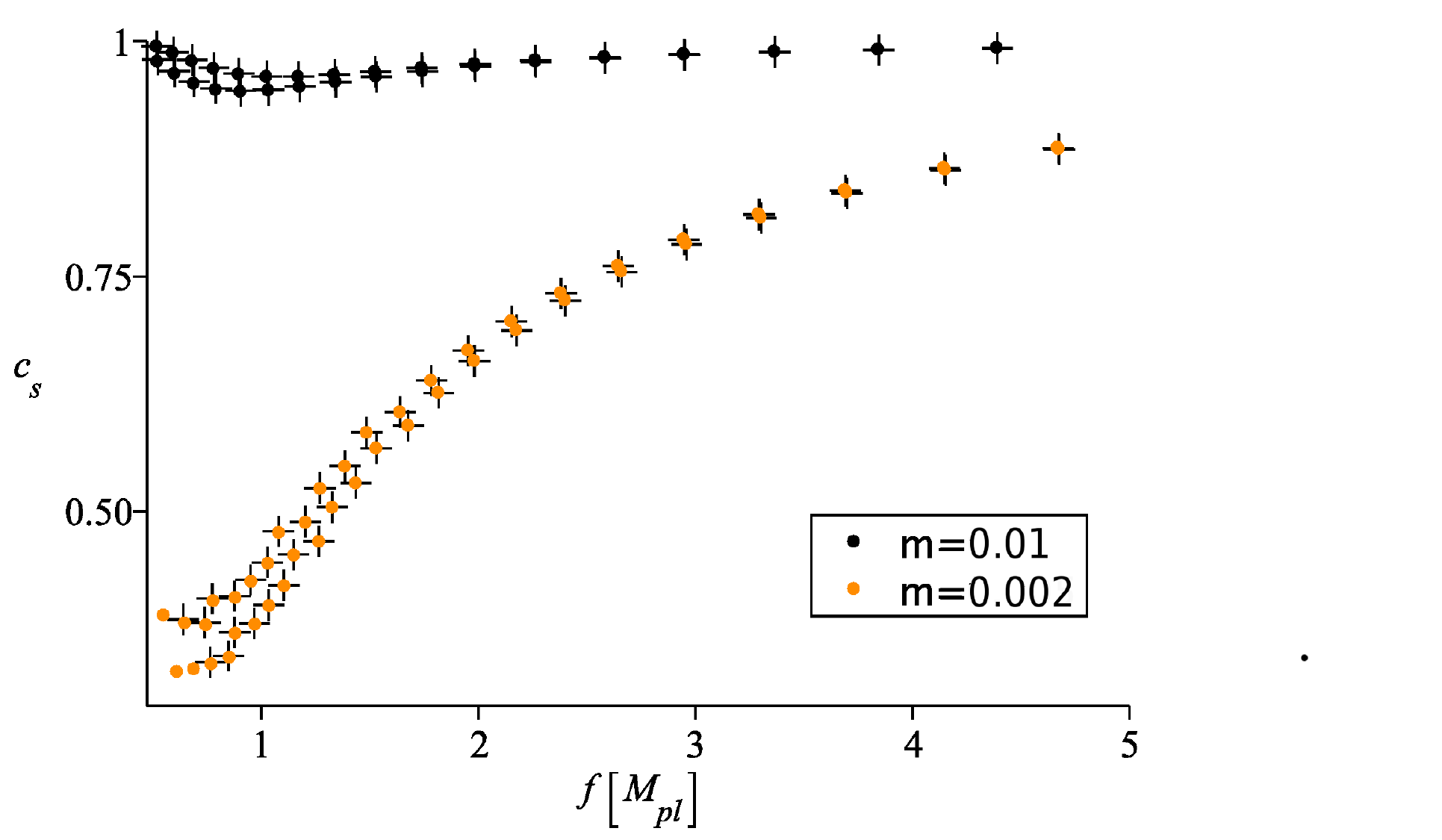}}
\caption{Speed of sound of the adiabatic fluctuations in the natural inflation model as a function of the instantaneous decay constant $f=\bar{\rho}/m$, for two values of $m$, $m=0.01$ and $m=0.002$. Each data point is double, representing the values for $c_s$ at N=50 (lower) and N=60 (upper). The dots are computed using the full two-field model while the crosses are obtained using the semi-analytical single field approximation.}
\label{fig:csNat}
\end{figure}

Finally, we consider the case for natural inflation \cite{Freese:1990rb}. The total Lagrangian is given by 

\be\label{eq:full_nat}
\mathcal{L}=\frac{1}{2}\partial_{\nu}\rho\partial^{\nu}\rho+\frac{1}{2}\rho^2\partial_{\nu}\theta\partial^{\nu}\theta-\frac{m_{\rho}^2}{2}\(\rho-\rho_0\)^2 -\Lambda^4\(1+\cos\(m\theta\)\) .
\ee
This two-field completion is consistent with the original motivation of the inflaton being a Nambu-Goldstone boson, as the additional field $\rho$ respects the $U(1)$ symmetry. If we assumed that the field $\rho$ acquires its v.e.v. at the minimum of the potential, and defining the canonical field as $\phi=\rho_0\theta$ we would get the standard potential, given by:
\be\label{eq:trunc_nat}
\mathcal{L}=\frac{1}{2}\partial_{\nu}\phi\partial^{\nu}\phi-\Lambda^4\(1+\cos\(\frac{\phi}{f_0} \)\).
\ee

\noindent where $f_0=\rho_{0}/m$ would be the so-called decay constant. Again, we will show that in general the dynamics of the radial field cannot be neglected - even at energies below the spontaneous symmetry breaking scale - so that (\ref{eq:trunc_nat}) is not a good description of (\ref{eq:full_nat}).\\
 While the natural - or axionic - potential is well motivated from the point of view of generic extensions of the Standard Model, the fact that decay constants larger than $M_{pl}$ are needed in order to achieve successful inflation is a problem from the point of view of the UV completion \cite{Banks:2003sx}. Attempts to construct effectively super planckian decay constants by considering several axions coupled together were first considered in \cite{Kim:2004rp,Dimopoulos:2005ac}.
  Here instead, we take a different route. We will show that the presence of a heavy degree of freedom can improve the situation, in the sense that the overlap in the $(n_s,r)$ plane between the predictions and the experimentally allowed region is greater. Since we cannot write an effective potential of the form (\ref{eq:trunc_nat}), it is not completely fair to talk about a decay constant in our model. However, if we consider an \emph{instantaneous} decay constant i.e. $f=\bar{\rho}/m$ (which is changing adiabatically as $\dot{f}/Hf=\dot{\bar{\rho}}/H\bar{\rho}\ll1$), we will show that indeed subplanckian values of $f$ could be consistent with the data. \\
 As in the case of the linear potential, it is a difficult task to find an effective potential, but we can compute the single field predictions by using the expressions for $\epsilon$, $\eta$ and $\Delta N$ as in eqs. (\ref{eq:sf_pred}) and $c_s$ and $s$ in eqs. (\ref{eq:cs}) and (\ref{eq:s}). We compare these predictions with the two-field model, in this case by considering two values for $m$, $m=\{0.01,0.002\}$, and find an excellent agreement. The results are shown in figures \ref{fig:Nat} and \ref{fig:csNat}.

Interestingly, as in the case of the linear and quadratic potential, the predictions move towards the best fit region. While the potential steepens due to the presence of the heavy field (so one might think the two-field model is disfavored with respect to the single field model), the reduced speed of sound and its variations causes the predictions to move towards the allowed experimental region.  We also notice that $0.8 M_{pl}<f<1 M_{pl}$ is consistent with Planck data for the case $m=0.002$. In the region where sub-planckian decay constants overlap with Planck contours, $|f^{eq}_{NL}|\sim 1$. Larger values of $|f^{eq}_{NL}|$ can be achieved by considering smaller values of $m$.  Whether the Lagrangian (\ref{eq:full_nat}) arises in some UV completion remains however an open question.

\section{Conclusion}\label{seccon}
Natural inflation and quadratic inflation are theoretically appealing scenarios but their single field realizations are in tension with the data. In general, one expects the single field description to be an EFT in which the effects of massive fields have been integrated out.
Crucially, while in some cases truncating the heavy field is a good approximation to construct the EFT, there is a regime in which the inflationary predictions can be very different from those of the single field truncated theory. This can happen even if the additional fields have masses much greater than the Hubble parameter. Considering that the heavy field tracks its instantaneous, adiabatic, ground state along the inflationary trajectory (which, on a turning trajectory, is displaced from the minimum of the potential) leads to modifications of the background evolution, as well as reducing the speed of sound of the light mode fluctuations. As we have shown, both effects are crucial in obtaining the correct predictions for where a model lies in the $(n_s,r)$ plane, as well as the expected level of non gaussianities. 

In this paper we have illustrated this idea in a very simple two-field embedding of various large field inflation models, in which the inflaton is approximately the phase of a complex field where a $U(1)$ symmetry is mildly broken. We find that this embedding can, in a weakly coupled regime, make  models that are in tension with the data, viable. In particular, the quadratic potential and the natural potential with subplanckian values of the decay constant are no longer disfavored. Although the effective potential \emph{steepens} in these examples, the effect of the speed of sound on the perturbations  dominates: while $\epsilon$ increases,  the tensor to scalar ratio goes \emph{down} due to the reduced speed of sound, $c_s$. Furthermore, adiabatic changes in $c_s$ along the trajectory can also modify of the spectral index. We have presented an analytic approximation which enables us to easily  calculate all the relevant observables

Finally, we should add that this phenomenology is not restricted to the particular embedding studied here i.e. a flat field-space metric and a separable potential. The essential characteristic is that inflation happens on a curved trajectory (with respect to the field-space metric) with a large, sustained mass hierarchy. The same approximate symmetry that protects the inflaton mass and slow roll parameters also keeps the radius of curvature and the mass of the heavy orthogonal direction approximately constant (for a recent discussion on adiabaticity and the slow roll conditions see \cite{Achucarro:2012yr} and references therein). In this regime the radial field can be integrated out, and the isocurvature perturbations decay quickly outside the horizon. The resulting effective theory for the perturbations has a reduced  speed of sound that changes slowly  along the trajectory. As shown here, the effects of this reduction can be very important in obtaining the correct predictions for inflationary observables.

\acknowledgments{We would like to thank Masahiro Kawasaki and Bin Hu for useful discussions. We are also specially grateful to Pablo Ortiz and Gonzalo Palma for long and enlightening discussions. VA would like to thank IPMU in Tokyo for hospitality during the completion of this work. This work was supported  by a Leiden Huygens Fellowship (VA), by an NWO - de Sitter Fellowship of the Netherlands Organization for Scientific Research (YW) and by the Netherlands Foundation for Fundamental Research on Matter (F.O.M.) under the program "Observing the Big Bang" (AA).  It was also supported by the Basque Government grant IT559-10, the Spanish Ministry of Science and Technology grant FPA2012-34456 and the Spanish Consolider-Iingenio 2010 program CPAN CDS2007-00042 (AA), }

\bibliography{PaperGen.bib}

\end{document}